\documentclass[%
 aip,
 amsmath,amssymb,
 reprint,%
]{revtex4-1}

\usepackage{graphicx}
\usepackage{dcolumn}
\usepackage{bm}

\usepackage[utf8]{inputenc}
\usepackage[T1]{fontenc}
\usepackage{mathptmx}
\usepackage{etoolbox}
\usepackage[version=3]{mhchem} 
\usepackage{tabularx}
\usepackage{graphicx}
\usepackage{booktabs}       
\usepackage{multirow}
\usepackage{makecell}
\DeclareUnicodeCharacter{2212}{-}

\makeatletter
\def\@email#1#2{%
 \endgroup
 \patchcmd{\titleblock@produce}
  {\frontmatter@RRAPformat}
  {\frontmatter@RRAPformat{\produce@RRAP{*#1\href{mailto:#2}{#2}}}\frontmatter@RRAPformat}
  {}{}
}%
\makeatother

\begin{document}

\preprint{AIP/123-QED}

\title[]{Beyond Independent Error Assumptions in Large GNN Atomistic Models}
\author{Janghoon Ock}
\author{Tian Tian}%
\author{John Kitchin}%
\author{Zachary Ulissi}%
 \email{zulissi@andrew.cmu.edu}
\affiliation{ 
Department of Chemical Engineering, Carnegie Mellon University 
}%


\date{\today}

\begin{abstract}
The practical applications of determining the relative difference in adsorption energies are extensive, such as identifying optimal catalysts, calculating reaction energies, and determining the lowest adsorption energy on a catalytic surface. Although Density Functional Theory (DFT) can effectively calculate relative values through systematic error cancellation, the accuracy of Graph Neural Networks (GNNs) in this regard remains uncertain. To investigate this issue, we analyzed approximately 483 million pairs of energy differences predicted by DFT and GNNs using the Open Catalyst 2020 - Dense dataset. Our analysis revealed that GNNs exhibit a correlated error that can be reduced through subtraction, thereby challenging the naive independent error assumption in GNN predictions and leading to more precise energy difference predictions. To assess the magnitude of error cancellation in chemically similar pairs, we introduced a new metric, the subgroup error cancellation ratio (SECR). Our findings suggest that state-of-the-art GNN models can achieve error reduction up to 77\% in these subgroups, comparable to the level of error cancellation observed with DFT. This significant error cancellation allows GNNs to achieve higher accuracy than individual adsorption energy predictions, which can otherwise suffer from amplified error due to random error propagation.

\end{abstract}
\maketitle



\section{Introduction}
\label{sec:intro}
Computational catalysis plays a crucial role in the development of more efficient and environmentally friendly chemical processes by predicting the behavior of catalysts. This allows researchers to identify the most promising catalyst candidates for a specific reaction, saving time and resources that would otherwise be spent on costly experimental screening \cite{CompCatal, Chen2021}. Adsorption energy, the energy required to adsorb a molecule onto a catalytic surface, is a key descriptor in computational catalysis \cite{Norskov2009, Yang2014}. The adsorption energy ($\Delta E_{\text{ads}}$) is obtained by subtracting the energy of the clean catalytic surface ($E_{\text{slab}}$) and the energy of the gas phase adsorbate molecule ($E_{\text{gas}}$) from the energy of the adsorbate-catalyst ($E_{\text{sys}}$). Adsorption energy is related to the reactivity of catalysts, so it can be used to estimate the reactivity of different catalysts \cite{Norskov2009, BEP, BEP2, Ulissi2017}.

\begin{equation}
    \centering
    \Delta E_{\text{ads}} = E_{\text{sys}} - E_{\text{slab}} - E_{\text{gas}}   \label{eq:Eads}
\end{equation}

\begin{figure}[hbt]
\centering
\includegraphics[width=0.485\textwidth]{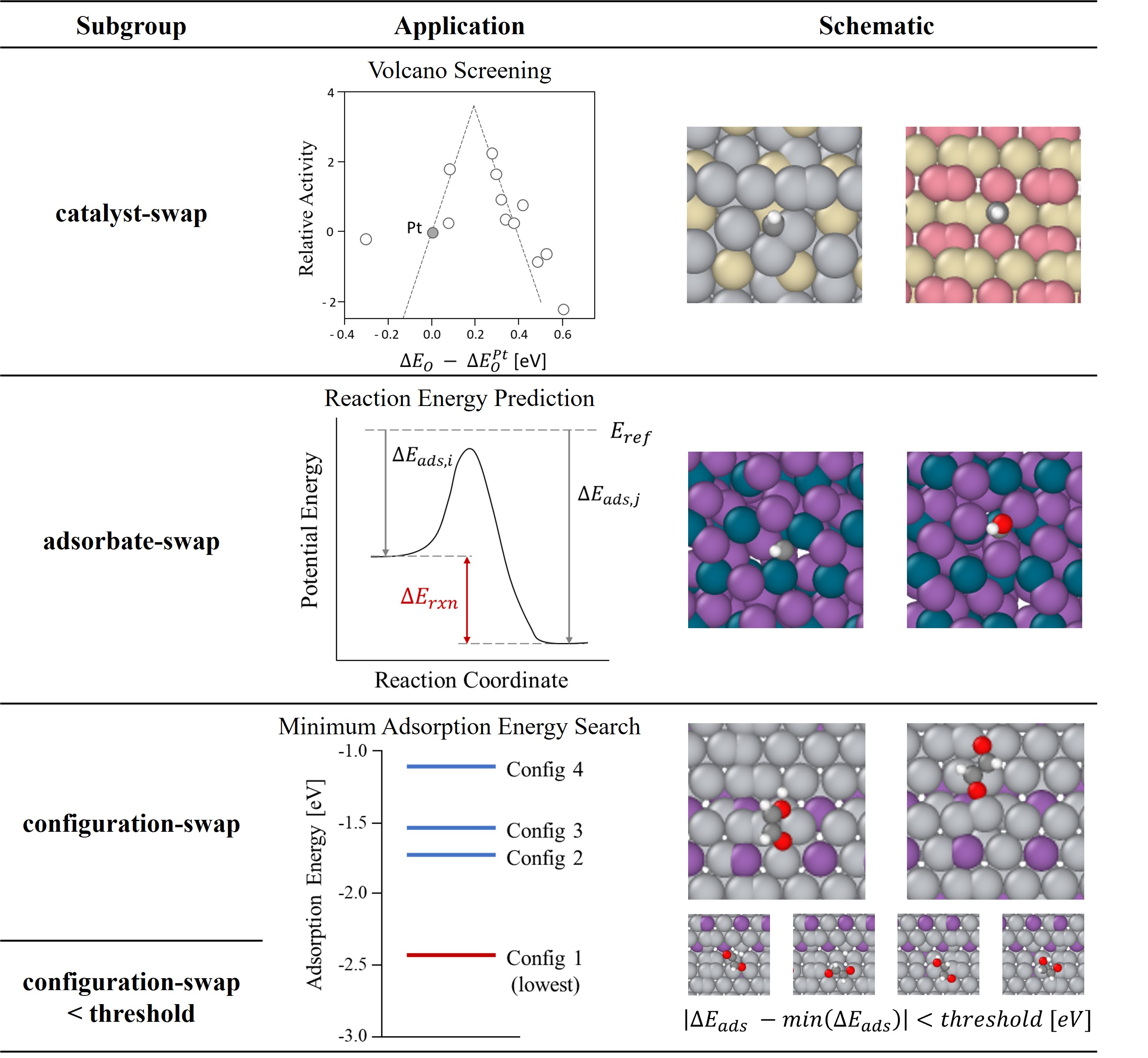} 
\caption{Demonstration of important use cases of the relative difference of adsorption energies in catalysis. The subgroups are named based on the element that is not shared by the pair. For example, the catalyst-swap subgroup represents pairs that do not share the catalytic surface, while the adsorbate-swap subgroup represents pairs that do not share the adsorbate molecule.}
\label{fig:subgroup}
\end{figure}

Descriptors in computational catalysis are more effective when predicting relative quantities than absolute energy values. In practical applications, the relative difference in adsorption energies ($\Delta \Delta E_{\text{ads}}$) between chemically similar systems is of particular interest, as described in Figure \ref{fig:subgroup}. For example, the adsorption energy of a specific surface can be obtained by identifying the minimum adsorption energy on the surface \cite{Ulissi2017, adsorbml}. In this case, the adsorption energies of different configurations should be compared. Additionally, a volcano plot compares the adsorption energies of a specific reactant or intermediate on various catalytic surfaces \cite{Li2019, Ooka2020, Cheng2008} to find the optimal catalyst, and reaction energies are calculated by comparing the adsorption energies of two atomic systems in a reaction pathway that have similar but different adsorbate molecules on the same surface \cite{Yang2014, Cheng2008}.

Density Functional Theory (DFT) is a widely used computational method for predicting adsorption energies, and it is more accurate at predicting relative values than absolute ones. This is because approximations in the DFT functionals often produce systematic errors that are correlated with the atomic structure of a system, which can be largely canceled out when comparing the adsorption energies of similar systems \cite{Collins2020, Philipp2020, Hautier2012}. This is known as error cancellation and can be quantified using the Bayesian Error Estimation with van der Waals functional (BEEF-vdW), which provides an ensemble of energy predictions rather than a single value \cite{beef_mortensen, beef_wellendorff}. The uncertainty from BEEF-vdW calculations reflects an error estimate associated with the approximation in the exchange-correlation (XC) functional, and it allows for a systematic evaluation of the uncertainty arising from variations in the XC functional \cite{beef_wellendorff, beef_mortensen, Venkat2019}. Error cancellation is particularly useful in computing the relative reactivity of catalysts, as the relative difference of reactivity has a smaller error range compared to the individual absolute reactivity \cite{Medford_science, Studt2021}. Therefore, in computational catalysis, the relative adsorption energies are practically more reliable quantities for catalyst screening and design \cite{Venkat2016}.

DFT is a reliable computational method based on quantum chemistry but can be computationally expensive, limiting the scale of studies. Machine Learning (ML) models trained with DFT results have been proposed as an alternative to reduce the computational cost \cite{MLforCat, OC20_intro, OC20}. Graph Neural Networks (GNNs) have emerged as a promising solution to accurately predict the interatomic energies and forces of atomic systems. In catalysis research, the Open Catalyst 2020 (OC20) dataset contains over 1.2 million DFT relaxations of adsorbate-catalyst systems that are used to train GNNs to replicate DFT calculations. With newer generations of GNNs, the accuracy of these models in predicting energies and forces has improved \cite{ocp_lb}. For adsorption energy prediction, the mean absolute error (MAE) is used as the primary error metric to evaluate the performance of GNN models \cite{OC20_intro, OC20}. The MAE has decreased as newer GNNs have been developed, and their accuracy has approached that of DFT calculations \cite{gemnet-oc, scn, ocp_lb}, which typically have an error range of 0.1 to 0.2 eV for adsorption energy prediction \cite{dft_accuracy}.

Although GNNs have achieved high accuracy in predicting individual adsorption energies, it is unclear whether they can predict relative differences with the same level of accuracy. Standard GNN models do not consider the correlation between nodes or graphs \cite{CGNN} and are not explicitly trained to predict relative differences. It is therefore yet to be determined if the ML-predicted values can benefit from the error cancellation observed in relative differences of DFT-predicted results \cite{Bartel2019, Bartel2020}. Despite the improved accuracy of newer GNN models in individual energy prediction, it is crucial to assess their ability to predict the relative difference of adsorption energies for practical applications such as creating volcano plots, calculating reaction energies, and identifying minimum energies on surfaces.

This study presents three primary contributions. Firstly, we demonstrate that errors in energy predictions made by GNNs trained on the OC20 dataset are correlated. This correlation enables the cancellation of errors when subtracting values. Secondly, we propose a metric for evaluating the performance of GNN models, which takes into account the cancellation of errors. This metric aims to develop more practical models that can accurately predict energy differences between catalyst systems. Finally, we propose enhancing error cancellation as a promising approach to improving the accuracy of energy difference predictions, rather than solely relying on improving individual energy prediction accuracy.

\section{Methods}
\label{sec:methods}
\subsection{Open Catalyst 2020 - Dense Dataset}
\label{subsec:sdataset}
We investigate the impact of structural factors, such as adsorbate molecules, catalytic surfaces, and adsorption configurations, on error cancellation using the Open Catalyst 2020 - Dense (OC20-Dense) dataset. This dataset was created to address the issue that the the adsorption energies in the OC20 dataset \cite{OC20_intro} may not correspond to the minimum energy of a specific adsorbate-catalyst combination. The OC20-Dense dataset collects 995 unique adsorbate-catalyst combinations from the OC20 dataset and densely enumerates initial configurations of adsorbates on surfaces, using both `heuristic' and `random' strategies \cite{adsorbml}. The heuristic strategy utilizes popular tools like CatKit \cite{catkit} and Pymatgen \cite{pymatgen}, while the random strategy places the adsorbate on randomly selected sites on the surface. DFT and ML relaxations are then conducted on all configurations, resulting in relaxed structures with distinct adsorption sites and orientations. The DFT relaxations use the revised Perdew-Burke-Emzerhof (RPBE) functional \cite{rpbe}, and the ML relaxations use multiple generations of GNNs, including SchNet \cite{schnet}, DimeNet++ \cite{dpp}, PaiNN \cite{painn}, GemNet-OC \cite{gemnet-oc}, and GemNet-OC-Large \cite{gemnet-oc}. In the GNN error analysis, the adsorption energies of the ML-relaxed structures are compared with those from the DFT-relaxed structures. 

In this research, the relaxation results on the configurations established by the heuristic approach are used as a baseline dataset since the DFT relaxations on them represent the common community baseline for evaluating multiple adsorption energies on the surfaces \cite{catkit, pymatgen}. OC20-Dense-Heuristic comprises 995 unique adsorbate-catalyst combinations spanning 76 adsorbates and 850 bulk catalysts, with an average of 31.2 configurations per combination \cite{adsorbml}. There are a total of 31,081 adsorbate-catalyst systems, resulting in approximately 483 million possible system pairs.

\begin{itemize}
\item \textbf{Adsorbate-catalyst combination}: A combination of adsorbate molecules and a catalytic surface, without specifying the adsorption configuration. 
 \item \textbf{Adsorbate-catalyst system}: A system comprising adsorbate molecules and a catalytic surface with a unique adsorption configuration, referred to as a `system' in this paper. Each combination in the OC20-Dense-Heuristics dataset has an average of 31 configurations, and each configuration represents an adsorbate-catalyst system.
 \item \textbf{System pair}: A pair of adsorbate-catalyst systems used to estimate their relative difference or similarity in this paper.
\end{itemize}

Additional DFT calculations with the BEEF-vdW functional are performed on top of the relaxed structures from RPBE functional in the OC20-Dense-Heuristics dataset to obtain energy ensembles for DFT error analysis. Due to the high computational cost, these calculations are only conducted on 1,529 systems, covering 201 adsorbate-catalyst combinations out of the 995 unique combinations in the OC20-Dense dataset.

\subsection{Subgrouping based on Similarity}
\label{subsec:subgrouping}
The chemical similarity between adsorbate-catalyst systems can be analyzed by considering two factors: similarity in the catalytic surface and similarity in the adsorbate molecule. To investigate these factors, three chemically-important subgroups can be created: \textbf{catalyst-swap (cat-swap)}, \textbf{adsorbate-swap (ads-swap)}, and \textbf{adsorption configuration-swap (conf-swap)} (see Figure \ref{fig:subgroup}). The cat-swap subgroup examines the impact of the catalytic surface on energy difference prediction and helps identify the optimal catalyst for a reaction using a Volcano plot. The ads-swap subgroup calculates reaction energies by comparing the adsorption energies of systems in the reaction pathway. The conf-swap subgroup identifies the minimum adsorption on a given surface by comparing adsorption energies with different sites and orientations. The systems that do not share any common elements are categorized as \textbf{all-swap}, representing the complement set of the three chemically similar subgroups.

\begin{itemize}
 \item \textbf{Cat-swap}: System pairs with the same surfaces, but different adsorbates.
 \item \textbf{Ads-swap}: System pairs with the same adsorbates, but different surfaces.
 \item \textbf{Conf-swap}: System pairs with the same surfaces and adsorbates, but different adsorption configurations.
 \item \textbf{All-swap}: System pairs do not share any common surfaces or adsorbates, which represents the complement of the three chemically similar subgroups.
\end{itemize}

To account for differences in adsorption energies that can arise from variations in adsorption configurations \cite{Liu2018, Bahamon2021, Deshpande2020}, even when catalytic surfaces and adsorbate molecules are identical, we can subgroup further based on similarity in adsorption energies. Similar adsorption configurations of the same adsorbate-surface combination typically have similar adsorption energies \cite{Gao2020}. We can further group pairs within the conf-swap subgroup based on their proximity to the surface's lowest adsorption energy and label them as either \textbf{conf-swap $<$ threshold} or \textbf{conf-swap within threshold}. The thresholds of energy difference and minimum adsorption energies are determined by the DFT calculation with RPBE. The thresholds are set as 1.0, 0.5, and 0.2 eV, and the average number of configurations in each subset is 16.7, 11.1, and 7.0 respectively. This subgrouping method helps identify the lowest adsorption energy on the surface through conf-swap analysis \cite{adsorbml}.

\begin{itemize}
 \item \textbf{Conf-swap $<$ threshold}: System pairs of which adsorption energies fall within a certain threshold from the lowest adsorption energy are selected from the entire conf-swap subgroup.  $\lvert \Delta E_{\text{ads}, i} - \Delta E_{\text{ads}, j} \rvert <$ thresholds (1.0, 0.5, 0.2 eV)
\end{itemize}

\begin{figure*}[ht] 
\centering
\includegraphics[width=0.85\textwidth]{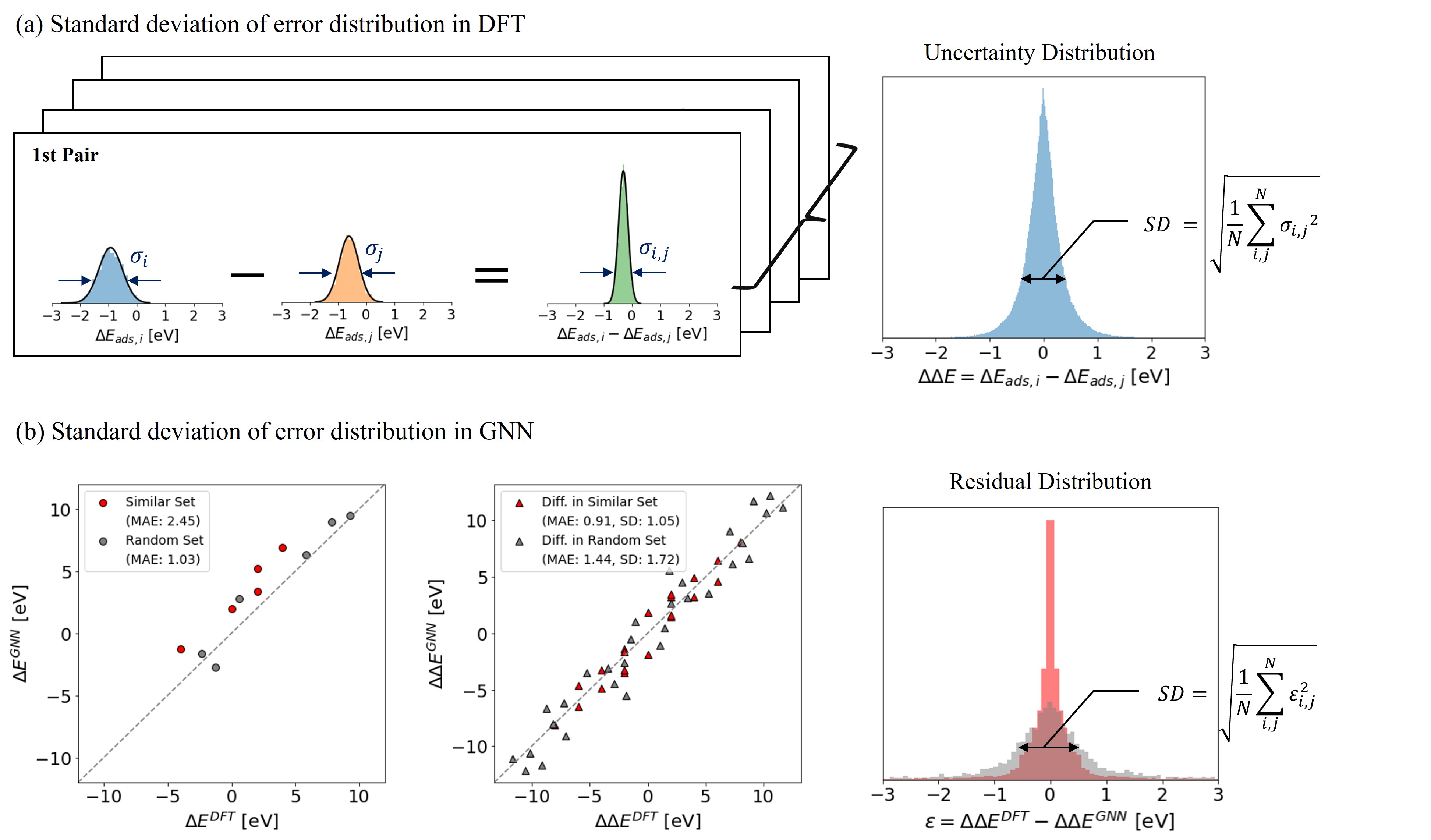} 
\caption{Standard deviation of the error distribution as a measure of the magnitude of the errors. (a) The standard deviation of the ensembles ($\sigma_i$, $\sigma_j$) for the BEEF-vdW functional can represent the error of individual calculations. Correlated ensembles result in a smaller $\sigma_{i,j}$ compared to the independent propagation of $\sigma_i$ and $\sigma_j$. (b) The residual distribution is obtained by comparing the energy differences calculated using the RPBE functional with those predicted by GNNs. As the mean of residuals ($\bar{\varepsilon}$) is zero in this case, the SD is equivalent to root-mean-square error (RMSE) of the energy difference prediction: $\sqrt{\frac{1}{N} \sum_{i,j}^{N} \left(\varepsilon - \bar{\varepsilon} \right)^2} = \sqrt{\frac{1}{N} \sum_{i,j}^{N} \varepsilon^2}$.}
\label{fig:framework}
\end{figure*}

\subsection{Metric for Error Cancellation}
\label{subsec:error-quantification}
The subgroup error cancellation ratio (SECR) is proposed as a metric that measures the magnitude of error cancellation in a certain subgroup. It is defined as the ratio of the standard deviation of error distribution (SD) in a subgroup to that of the entire energy difference predictions. A lower SECR value indicates a greater degree of error cancellation within the subgroup. By analyzing the SECR of various subgroups, we can gain insight into the extent of error cancellation and identify the factors that contribute most to this phenomenon. As previously mentioned, a high correlation between errors leads to strong error cancellation and a reduction in the overall error distribution. Therefore, the magnitude of the reduction in the error distribution within a specific subgroup can be used as an indicator of a strong error correlation.

\begin{equation}
    \centering
    \text{SECR [\%]} = 1-\frac{\text{SD in subgroup}}{\text{SD in total dataset}}
    \label{eq:SECR_dft}
\end{equation}

The BEEF-vdW functional is a DFT method that estimates the uncertainty in DFT-calculated energies by generating an ensemble of energies \cite{beef_mortensen, beef_wellendorff}. The uncertainty ensemble of energy differences is obtained by subtracting the energy ensembles, as described in Figure \ref{fig:framework}a. The uncertainty distribution of a specific subgroup is equivalent to the accumulation of all the energy difference ensembles within the subgroup. Therefore, the standard deviation of the resulting uncertainty distribution (SD) is equivalent to the average of the standard deviations of the energy difference ensembles ($\sigma_{i,j}$) within the subgroup. Further details are provided in the Supplementary Information (SI).

The standard deviation of the residual distribution (SD) is used as an estimate of the error variance of the GNN prediction for a specific subgroup, as described in Figure \ref{fig:framework}b. In the example case shown in Figure \ref{fig:framework}b, a certain subset may exhibit a smaller error in the energy difference, despite a larger error in the individual adsorption energy prediction. This is because the errors in the individual energy prediction cancel out during subtraction. Thus, strong error cancellation can be a promising approach to reduce the error in the energy difference prediction.

\subsection{Correct Sign Prediction in Energy Difference}
\label{subsec:csr}
Inaccurate predictions, especially when the sign of the energy difference is wrong, can lead to opposite trends in reaction activity and ultimately the failure of the screening process \cite{screen, Bartel2020}. The challenge of determining the correct sign of the energy difference is heightened when comparing similar adsorption energies. Therefore, the capability of capturing the correct sign of energy difference within a certain margin is a good indicator of the energy difference prediction's quality. We propose using the marginal correct sign ratio (MCSR) as a quantitative measure for this purpose. The MCSR is defined as the ratio of the number of correct sign predictions to the total number of data points within a certain margin value in DFT-calculated energy differences. In this paper, the margin is set as 0.2 eV so as to evaluate the model performance for determining the subtle difference.

\begin{equation}
    \centering
    \text{MCSR [\%]} = \frac{\sum\limits_{i,j}^N \left[ \Delta \Delta E_{ij}^{\text{DFT}} \Delta \Delta E_{ij}^{\text{GNN}} > 0 \right] \left[\lvert \Delta \Delta E_{ij}^{\text{DFT}} \rvert < \text{margin} \right]}{\sum\limits_{i,j}^N \left[\lvert \Delta \Delta E_{ij}^{\text{DFT}} \rvert < \text{margin} \right] }
    \label{eq:mcsr}
\end{equation}

\section{Results \& Discussion}
\label{sec:results_discussion}

\subsection{Data Fraction and Energy Difference of Subgroups}
\label{subsec:fraction_range}



\begin{table} 
\caption{Distribution of various types of energy differences in the OC20-Dense-Heuristics dataset.}
\label{table:data-and-range}
\resizebox{0.48\textwidth}{!}{%
\begin{tabular}{@{}lcccc@{}}
\toprule
 \multicolumn{2}{c}{\textbf{Subgroup}} & \textbf{Data Proportion \footnote{the ratio of pairs in a subgroup to the total system pairs.} [\%]} &  \textbf{Avg. Energy Diff. \footnote{$\frac{1}{N}\sum_{i}^N \lvert \Delta \Delta E_{ads, i}\rvert$, where $N$ is the number of pairs in a subgroup.} [eV]}  \\

\midrule
All-swap           &          & 93.78 & 2.23   \\ \midrule
Catalyst-swap      &          & 6.01  & 1.43   \\
Adsorbate-swap     &          & 0.03  & 2.04   \\
\multirow{4}{*}{Configuration-swap} & (all)            & 0.22  & 0.83  \\
                                    &\textless 1.0 eV  & 0.06  & 0.28  \\
                                    &\textless 0.5 eV  & 0.03  & 0.14  \\ 
                                    &\textless 0.2 eV  & 0.01  & 0.05  \\ \midrule
Total              &          & 100   & 2.18    \\ \bottomrule
\end{tabular}

}
\end{table}

The small representation of certain subgroups in the OC20-Dense-Heuristics dataset can lead to a lack of focus on the prediction accuracy for those subgroups. In practice, it is often necessary to compare catalyst systems that have common components to extract meaningful information, such as reaction energy and minimum adsorption energy on the surface, as depicted in Figure \ref{fig:subgroup}. Despite this, less than 7\% of the pairs in the dataset share a common catalyst or adsorbate, as shown in Table \ref{table:data-and-range}. Furthermore, the data portions for ads-swap and conf-swap are very scarce, accounting for only 0.03\% and 0.22\%, respectively, of the total pairs. This means that the prediction accuracy for these practically meaningful subgroups may be overlooked when evaluating the performance of a GNN model over the entire energy difference.

Accurately predicting energy differences in subgroups with high similarity can be challenging, as the adsorption energies of chemically relevant pairs tend to be similar.For example, the conf-swap within 0.5 eV exhibits an average energy difference of 0.14 eV, and the config-swap within 0.2 eV shows 0.05 eV, as calculated using the RPBE functional. These subtle energy differences are even smaller than the MAE of adsorption energy predictions made by state-of-the-art GNNs \cite{ocp_lb}, making it difficult to distinguish system pairs in this subgroup. 

\begin{figure*}[hbt]
\includegraphics[width=0.75\textwidth]{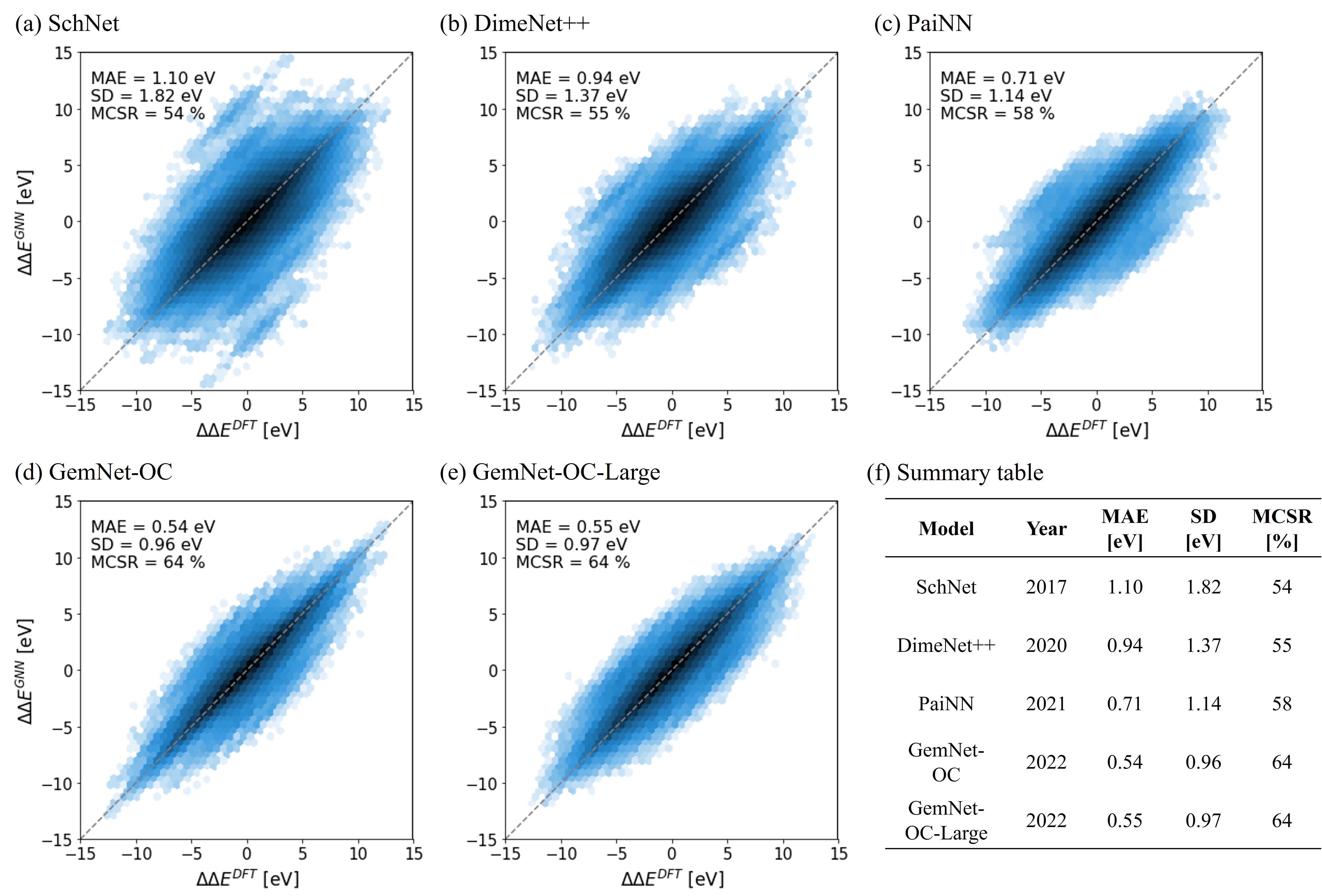} 
\caption{Parity plots comparing the energy difference predicted by GNNs - (a) SchNet, (b) DimeNet++, (c) PaiNN, (d) GemNet-OC, and (e) GemNet-OC-Large - with the energy difference calculated by DFT. (f) A summary table for energy difference predictions.}
\label{fig:parity}
\end{figure*}

GNNs are not trained to explicitly predict the relative difference between adsorption energies, and thus the energy difference is calculated by subtracting the ML-predicted adsorption energies. To evaluate the accuracy of GNN predictions, the energy differences obtained from both DFT and GNNs are compared, as demonstrated in Figure \ref{fig:parity}. SchNet, DimeNet++, PaiNN, and GemNet-OC were published in 2017, 2020, 2021, and 2022, respectively. The release of newer models has led to a decrease in MAE for the energy difference prediction of the OC20-Dense-Heuristics.

\subsection{Cancellation of Errors across Subgroups of OC20-Dense}
\label{subsec:error_cancel}

The errors in DFT serve as a benchmark for evaluating the errors in GNNs, as the DFT values are utilized as training labels for GNNs. The uncertainty of DFT calculations is reduced across chemically similar subgroups, such as cat-swap, ads-swap, and conf-swap, through error cancellation, as described in Table \ref{table:SECR}. Since the errors in DFT calculations are correlated, they can be reduced by subtracting them from one another. The all-swap subgroup, where the pairs do not share any common adsorbate or surface, shows negative SECR, meaning that the errors are amplified rather than reduced. The uncertainty distribution of the energy differences between chemically similar pairs is therefore sharper than that of chemically irrelevant pairs (see SI). The order of SECR for DFT follows the order: conf-swap $<$ cat-swap $\simeq$ ads-swap $<$ all-swap, with the greatest SECR being observed in the conf-swap $<$ 0.2 and 0.5 eV. The uncertainty distribution in the conf-swap subgroup is significantly reduced compared to the other two chemically similar subgroups, showing approximately 50\% higher SECR of the cat-swap and ads-swap subgroups.

\begin{table}[ht]
\caption{Subgroup error cancellation ratio (SECR) of DFT and GNN across subgroups.}
\label{table:SECR}
\resizebox{0.48\textwidth}{!}{%
\begin{tabular}{@{}lcccccccc}
\toprule
 &  \multicolumn{6}{c}{\textbf{Subgroup Error Cancellation Ratio [\%] (higher is better)}} \\ \midrule
\multirow{2}{*}{\textbf{Model}} &  \multirow{2}{*}{\textbf{All-swap}} & \multirow{2}{*}{\textbf{\thead{Catalyst \\ -swap}}} & \multirow{2}{*}{\textbf{\thead{Adsorbate \\ -swap}}} & \multicolumn{4}{c}{\textbf{Configuration-swap}} \\

& & & & \textbf{(all)} & \textbf{\textless 1.0 eV} & \textbf{\textless 0.5 eV} & \textbf{\textless 0.2 eV} \\
\midrule
SchNet          & 0.2  & -3  & -19 & 22  & 33  & 35  & 35 \\
DimeNet++       & -0.8 & 13  & 10  & 20  & 62  & 70  & 69 \\
PaiNN           & -0.6 & 9   & 12  & 19  & 60  & 68  & 66 \\
GemNet-OC       & -0.6 & 10  & 2   & 14  & 62  & 69  & 68 \\
GemNet-OC-Large & -0.7 & 12  & 5   & 15  & 64  & 71  & 71 \\ 
\midrule
BEEF-vdW        & -3 & 42  & 42  & 59 & 65  & 72  & 77 \\
\bottomrule
\end{tabular}
}

\end{table}
The results of the GNN predictions demonstrate a positive SECR across various subgroups, including conf-swap, cat-swap, and ads-swap. This outcome constitutes the first demonstration that the errors in GNNs for large-scale atomistic simulations are also not independent, and can be partially canceled out when taking differences of energies. SchNet, one of the earliest GNN models, shows a positive SECR exclusively in the conf-swap subgroup. Other GNN models, meanwhile, showed a reduction in error distribution for all chemically similar subgroups. Furthermore, GNNs tend to exhibit higher SECR for conf-swap subgroups compared to cat-swap and ads-swap, and higher SECR as the threshold for conf-swap within thresholds decreases, aligning with observations in DFT.

\begin{table}[t]
\caption{Comparison of the standard deviation of error distribution (SD) across subgroups and GNN models.}
\label{table:SD}
\resizebox{0.48\textwidth}{!}{%
\begin{tabular}{@{}lccccccccc}
\toprule
&  \multicolumn{8}{c}{\textbf{Standard Deviation of Error Distribution [eV] (lower is better)}} \\ \midrule 
 &  \multicolumn{2}{c}{\textbf{Total \footnote{The results for the entire OC20-Dense-Heuristics dataset.}}} & \multicolumn{7}{c}{\textbf{Subgroup}}   \\
\cmidrule(lr){2-3} \cmidrule(lr){4-10} 

\multirow{2}{*}{\textbf{Model}} & \multirow{2}{*}{\textbf{$\Delta E_{\text{ads}}$}} & \multirow{2}{*}{\textbf{$\Delta \Delta E_{\text{ads}}$}} & \multirow{2}{*}{\textbf{All-swap}} & \multirow{2}{*}{\textbf{\thead{Catalyst \\ -swap}}} & \multirow{2}{*}{\textbf{\thead{Adsorbate \\ -swap}}} & \multicolumn{4}{c}{\textbf{Configuration-swap}}\\

& & & & & & \textbf{(all)} & \textbf{\textless 1.0 eV} & \textbf{\textless 0.5 eV} & \textbf{\textless 0.2 eV}  \\

\midrule
SchNet          & 1.29 & 1.82 & 1.82 & 1.88 & 2.17 & 1.43 & 1.22 & 1.18 & 1.18   \\
DimeNet++       & 0.97 & 1.37 & 1.38 & 1.20 & 1.24 & 1.10 & 0.52 & 0.41 & 0.42  \\
PaiNN           & 0.80 & 1.14 & 1.15 & 1.04 & 1.01 & 0.82 & 0.45 & 0.37 & 0.39  \\
GemNet-OC       & 0.68 & 0.96 & 0.96 & 0.86 & 0.94 & 0.82 & 0.36 & 0.30 & 0.31  \\
GemNet-OC-Large & 0.68 & 0.97 & 0.98 & 0.86 & 0.92 & 0.82 & 0.35 & 0.28 & 0.28  \\
\midrule
BEEF-vdW        & 0.38 & 0.42 & 0.44 & 0.25 & 0.25 & 0.17 & 0.15 & 0.12 & 0.10  \\
\bottomrule
\end{tabular}
}
\end{table}

\subsection{Magnitude of Error Cancellation in DFT and GNNs}
\label{subsec:magnitude}

Since the error distribution of DFT and GNN are defined differently in the SECR, it is not valid to directly compare the SECR of DFT and GNN in Table \ref{table:SECR}. Instead, to determine whether error cancellation in GNNs is comparable to DFT, we must estimate if the error cancellation can effectively reduce the errors in subgroups of energy differences to the level of errors from individual adsorption energy predictions. Table \ref{table:SD} shows that for DFT, the overall SD of the uncertainty distribution (0.42 eV) is higher than that for individual energy predictions (0.38 eV), but all subgroups except for all-swap have a lower SD, ranging from 0.10 to 0.25 eV. This is consistent with the known fact that relative DFT results for similar atomic systems are more reliable than absolute individual values. 

In contrast, a similar level of error to individual energy prediction is only seen in subgroups of conf-swap within thresholds. For all GNNs, the SDs for all-swap, cat-swap, ads-swap, and the entire conf-swap are still larger than their SDs of individual energy prediction. This suggests that error cancellation in GNNs is not as strong as in DFT, where the accuracy improvement is seen in cat-swap, ads-swap, and conf-swap. Therefore, for GNNs, it may not be possible to achieve the same level of accuracy as the individual energy prediction for tasks such as calculating reaction energy (represented by ads-swap) or comparing reactivity between different catalyst surfaces (represented by cat-swap).

\begin{table}[hb]
\caption{Comparison of the marginal correct sign ratio (MCSR) across subgroups and GNN models with a margin of 0.2 eV. Conf-swap within threshold subgroups exhibits improved accuracy in predicting correct signs across all cases, with larger MCSRs providing evidence of this increased capability.}
\label{table:MCSR}
\resizebox{0.48\textwidth}{!}{%
\begin{tabular}{@{}lccccccccc}
\toprule
 &  \multicolumn{7}{c}{\textbf{Marginal Correct Sign Ratio [\%] (higher is better)}} \\ \midrule
\multirow{2}{*}{\textbf{Model}} &  \multirow{2}{*}{\textbf{Total}} &\multirow{2}{*}{\textbf{All-swap}} & \multirow{2}{*}{\textbf{\thead{Catalyst \\ -swap}}} & \multirow{2}{*}{\textbf{\thead{Adsorbate \\ -swap}}} & \multicolumn{4}{c}{\textbf{Configuration-swap}} \\

& & & & & \textbf{(all)} & \textbf{\textless 1.0 eV} & \textbf{\textless 0.5 eV} & \textbf{\textless 0.2 eV} \\
\midrule
SchNet          & 54 & 54 & 54 & 54 & 54 & 55 & 56 & 53 \\
DimeNet++       & 55 & 55 & 55 & 54 & 56 & 58 & 59 & 55 \\
PaiNN           & 58 & 58 & 58 & 58 & 60 & 62 & 63 & 59 \\
GemNet-OC       & 64 & 64 & 64 & 64 & 66 & 69 & 70 & 66 \\
GemNet-OC-Large & 64 & 64 & 64 & 68 & 67 & 70 & 71 & 67 \\ 
\bottomrule
\end{tabular}
}
\end{table}

The strong error cancellation observed for conf-swap within certain thresholds in the GNN results enhances their ability to accurately determine subtle energy differences. The MCSR provides valuable insights into the ability of GNNs to detect the correct sign of small energy differences, particularly those below 0.2 eV. As shown in Table \ref{table:MCSR}, all GNN models exhibit higher MCSR values for conf-swap within thresholds, compared to other subgroups or entire pairs, despite those subgroups having the smallest energy differences. This outcome is consistent with the findings presented in Table \ref{table:SD}, which demonstrates that the error distributions of conf-swap within thresholds are sharper than those of the entire energy difference. Error cancellation improves the accuracy of subtle energy difference determination, enabling GNNs to correctly identify the sign of subtle energy differences.

\subsection{Trends in SECR and MAE across GNNs}
\label{subsec:mae_vs_secr}

\begin{figure}[hbt]
\centering
  \includegraphics[width=.9\linewidth]{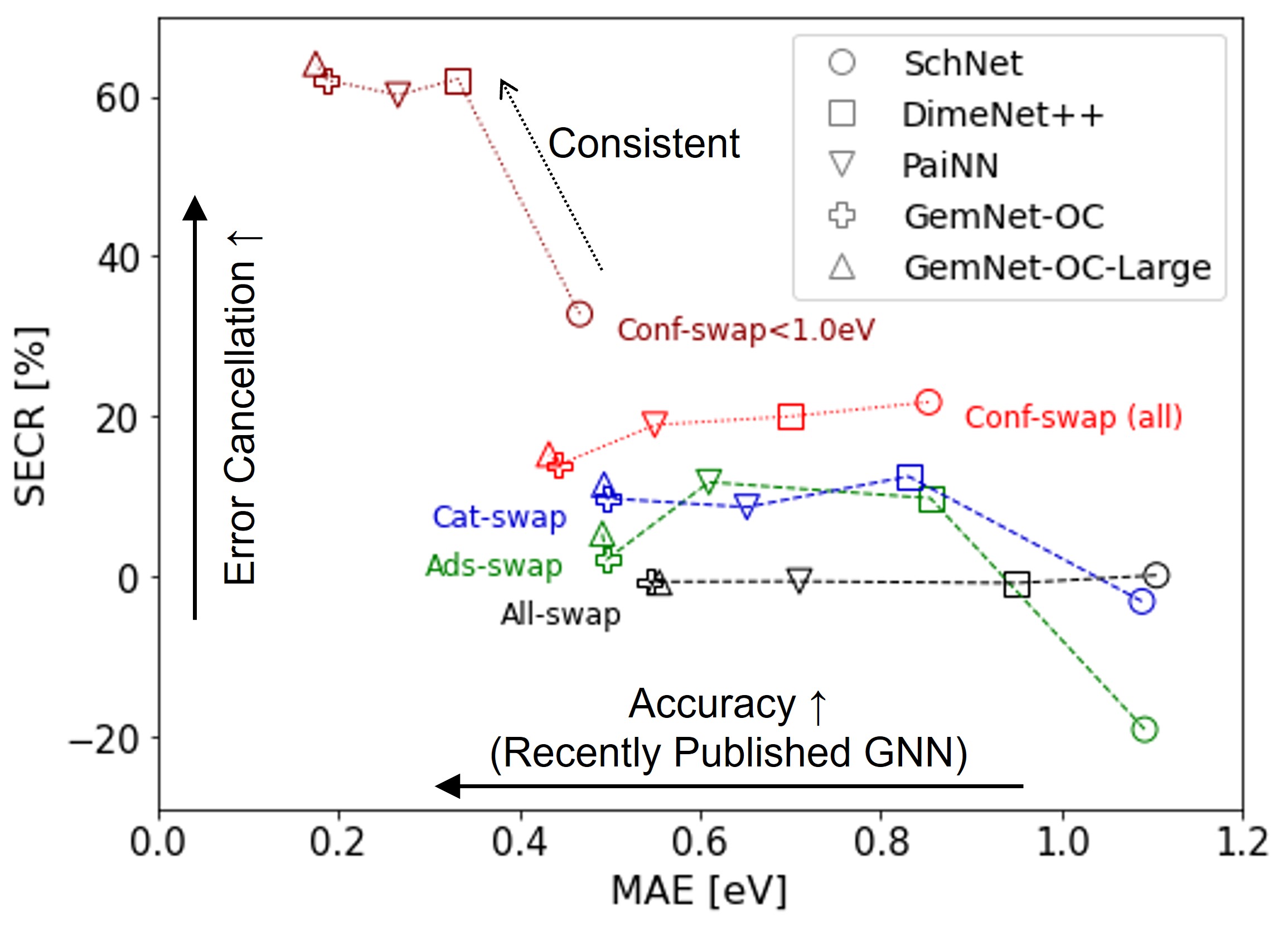}
\caption{Comparison of SECR against MAE. MAE is calculated from the energy difference prediction over the OC20-Dense-Heuristics dataset. If the recent GNN models show a consistent improvement in SECR as the MAE decreases, we would anticipate observing a gradual shift in each subgroup plot from the bottom right to the upper left (indicated by the dotted arrow).}
\label{fig:mae_vs_secr}
\end{figure}

Recent versions of GNNs have shown improved performance over the benchmark OC20 dataset, as noted in previous studies \cite{ocp_lb}. This improvement extends to the prediction of energy differences on the OC20-Dense-Heuristics dataset, as shown in Figure \ref{fig:parity}. However, Figure \ref{fig:mae_vs_secr} reveals that the trend of decreasing MAE does not consistently correspond with the trend of SECR for chemically similar subgroups. While there is a significant increase in SECR for GNN models after SchNet, subsequent models such as DimeNet++, PaiNN, GemNet-OC, and GemNet-OC-Large do not differ significantly in SECR. Notably, GemNet-OC and GemNet-OC-Large, which are state-of-the-art models, exhibit smaller SECR in the ads-swap, cat-swap, and entire conf-swap subgroups than DimeNet++, despite their lower MAE. The magnitude of error cancellation does not align with the progression of GNNs, indicating that additional reduction in MAE in energy difference prediction is possible by adopting suitable error cancellation methods.

\subsection{Correlation of Latent Space Distances with Error Cancellation}
\label{subsec:embedding}

\begin{figure*}[hbt]
\centering
  \includegraphics[width=.75\linewidth]{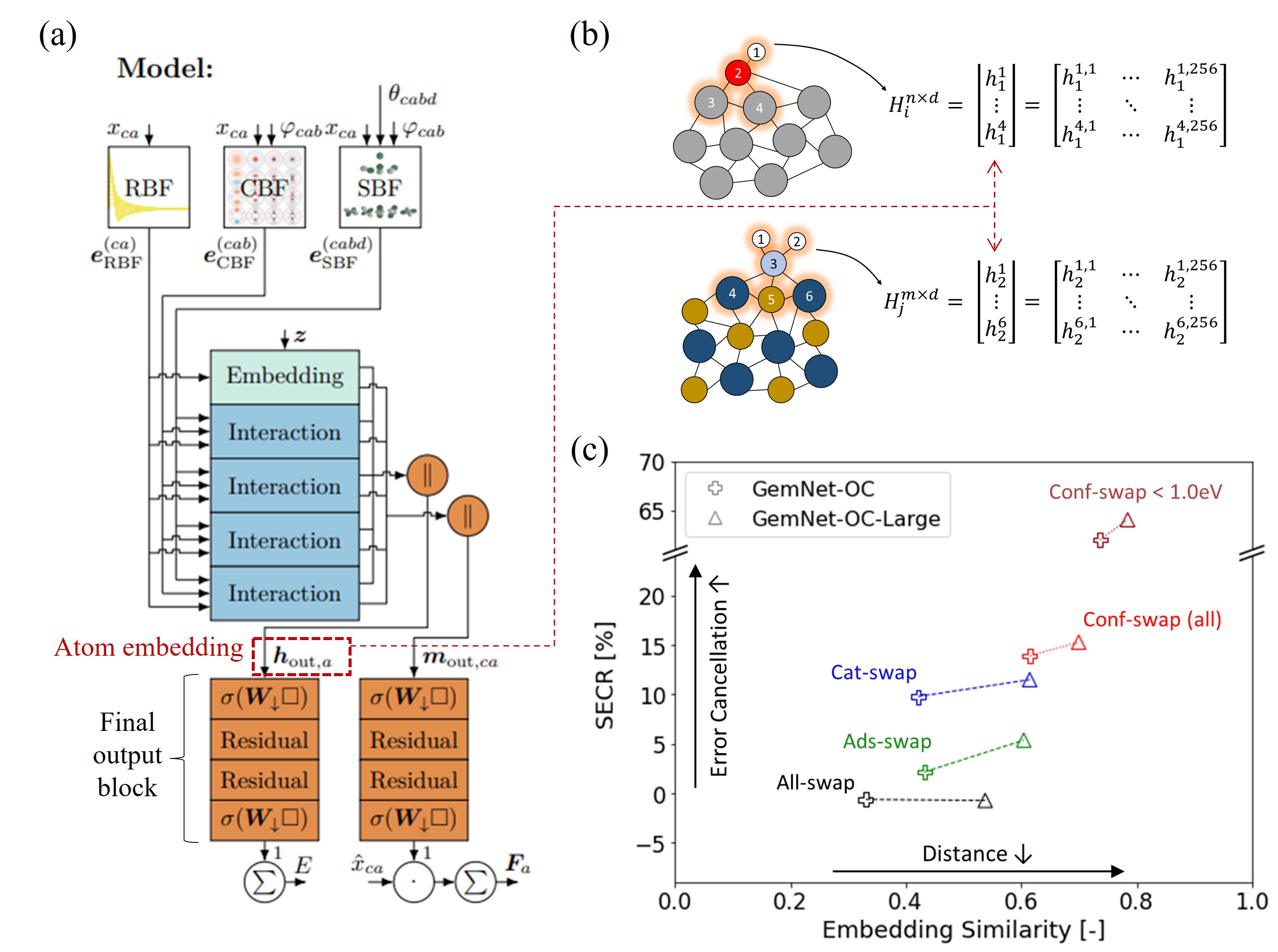}
\caption{Process of atom embedding analysis. (a) The model architecture of GemNet-OC \cite{gemnet-oc} is shown, where $h_{\text{out, a}}$ represents the atom embedding of atom \textit{a}. Atom embeddings of the adsorbate and its nearest neighbors on the catalytic surface (ads-NN) are collected before the final output block (orange). (b) The atom embeddings of ads-NN are concatenated to construct a 2-dimensional matrix of size $n$ by $d$, where $n$ is the number of atoms and $d$ is the embedding dimension (256 for GemNet-OC and GemNet-OC-Large). The glowing atoms in this figure represent the ads-NN. (c) The SECR is plotted against the similarity between the atom embeddings of ads-NN. High similarity indicates close proximity in the latent space.}
\label{fig:embedding}
\end{figure*}

The latent space of GNNs captures energy prediction characteristics through graph-based methods, transforming molecules into numerical embeddings that include atom-specific information. These embeddings are iteratively updated through message passing between neighboring atoms \cite{schnet, dimenet, dpp, painn, gemnet, gemnet-oc}. Figure \ref{fig:embedding}a shows the architecture of GemNet-OC, where the final atom embeddings ($h_{\text{out}}$) are used in the output block to predict the energy of each atom node ($e$) and then summed to obtain the predicted energy of the adsorbate-catalyst system ($E)$. The distance between final atom embeddings in the latent space indicates system similarity \cite{Yilin2022}.

In this study, we focus on the atom embeddings of the adsorbate and its nearest neighbors on the catalytic surface (ads-NN) for similarity analysis. This selection is necessary to prevent the similarity analysis from being biased by the number of atoms. For instance, comparing the entire set of atoms may lead to lower embedding similarity for cat-swap compared to conf-swap and ads-swap, as the atoms of the catalyst, which account for approximately 90\% of the system, are not shared in cat-swap pairs.

A 2-dimensional matrix ($H^{n \times d}$) is obtained by selectively concatenating the atom embeddings from the ads-NN atoms before the final output block, as depicted in Figure \ref{fig:embedding}b. The size of the matrix depends on the number of atoms ($n$) and the size of the atom embedding vectors ($d$), which can vary across GNN models, e.g., SchNet uses an embedding size of 1,024, DimeNet++ uses 192, PaiNN uses 512, and both GemNet-OC and GemNet-OC-Large use 256. The RV coefficient, a multivariate generalization of the squared Pearson correlation coefficient \cite{rv_coeff, rv_coeff2}, is used to measure the similarity of embedding matrices on a scale from 0 to 1. This coefficient enables comparisons between matrices of different sizes, as each catalyst system contains a distinct number of atoms in ads-NN. Further details are provided in the SI.


By comparing GemNet-OC and GemNet-OC-Large, we can isolate the impact of model training on embedding similarity since both models have the same model architecture. GemNet-OC-Large is trained on a larger dataset that includes additional molecular dynamics simulations on a random subset of the original OC20 (OC-MD), whereas GemNet-OC is trained on a regular OC20 dataset. Although both models exhibit comparable accuracy in adsorption energy and energy difference predictions (see Tables \ref{table:SECR} and \ref{table:SD}), GemNet-OC-Large shows higher SECR across all chemically similar subgroups compared to GemNet-OC. Figure \ref{fig:embedding} illustrates the relationship between ads-NN embedding similarity and SECR. For chemically similar pairs, GemNet-OC-Large generates more similar atom embeddings and shows stronger error cancellation compared to GemNet-OC. This finding suggests that training GNN models to produce highly correlated ads-NN atom embeddings for similar structures can lead to stronger error cancellation, without the need for modifications to the model architecture.

\section{Conclusion}
\label{sec:conclusion}

This study demonstrates that GNNs trained on the OC20 dataset for large-scale atomistic simulations exhibit correlated errors resulting in error cancellation, similar to those observed in DFT calculations. Our analysis introduces a metric for error cancellation, enabling the comparison of error cancellation magnitudes across subgroups. Notably, GNNs exhibit a reduced error distribution when predicting energy differences for chemically similar adsorbate-catalyst systems.

The ability of GNNs to accurately determine the relative difference in adsorption energies for chemically similar catalyst systems is crucial for practical applications, including volcano screening for optimal catalysts, reaction energy calculations, and identifying the optimal adsorption configuration. The substantial error cancellation observed in the conf-swap within threshold subgroups demonstrates the GNNs' ability to distinguish subtle differences in adsorption energies close to their minimum. This aligns with previous research utilizing GNNs to determine the minimum adsorption energy. Our study shows that GNNs are particularly good at this task, as the errors in the systems with adsorption energies close to the minimum exhibit high cancellability.

Despite exhibiting error cancellation through correlated errors, GNNs fall short in comparison to DFT in terms of error correlation and cancellation. The error distribution of energy difference predictions made by GNNs is generally larger than that of individual energy predictions, indicating that the current GNNs do not show the same level of error cancellation as DFT. Additionally, the level of error cancellation does not consistently improve with the development of newer GNNs that have lower MAE. Therefore, enhancing error cancellation can be a promising approach for improving the accuracy of energy difference predictions beyond individual adsorption energy predictions. In this regard, error cancellation for chemically similar system pairs should be considered a critical metric for assessing the performance of GNNs in large-scale atomistic simulations.

\section*{Supplementary Information}
\label{sec:si}
The supplementary material includes: Error cancellation in BEEF-vdW; error distributions in GNN predictions; contribution of atom embeddings from the adsorbate and its nearest neighbor atoms on the surface in energy prediction; embedding similarity analysis using the RV coefficient

\begin{acknowledgments}
\label{sec:acknowledgements}
The authors thank Aini Palizhati, Muhammed Shuaibi, and Janice Lan for their generous sharing of valuable data and their insightful discussions, which significantly contributed to the success of this study. 
\end{acknowledgments}

\section*{Author declarations}
\subsection*{Conflict of Interest}
The authors have no conflicts to disclose

\subsection*{Author Contributions}
Janghoon Ock: Conceptualization (equal); Data curation (equal); Formal analysis (equal); Investigation (equal); Methodology (equal); Visualization (equal); Writing – original draft (equal); Writing – review \& editing (equal). Tian Tian: Conceptualization (equal); Data curation (equal); Methodology (equal); Validation (equal); Investigation (equal). John Kitchin: Conceptualization (equal); Methodology (equal); Writing – original draft (equal); Writing – review \& editing (equal). Zachary Ulissi: Conceptualization (equal); Funding acquisition (lead); Supervision (lead); Project administration (lead); Writing – original draft (equal); Writing – review \& editing (equal).

\section*{Data Availability Statement}
The Open Catalyst 2020 - Dense dataset is openly available at \url{https://github.com/Open-Catalyst-Project/AdsorbML}.



\bibliography{reference}
\end{document}


\tableofcontents

\newpage

\section{Error cancellation in BEEF-vdW}
\label{sec:s1}
\begin{figure}[h] 
\begin{subfigure}{.9\textwidth}
  \centering
  \caption[]{\small All-swap}
  \includegraphics[width=.99\textwidth]{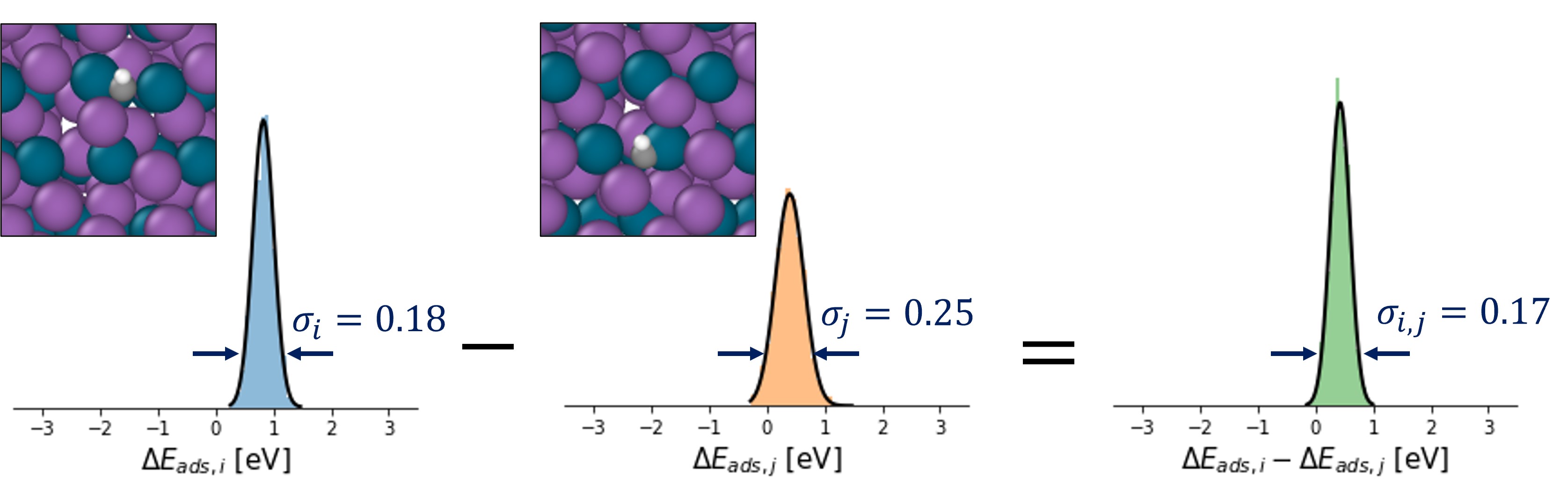}
  \label{fig:s1_a}
\end{subfigure}
\vskip\baselineskip
\begin{subfigure}{.9\textwidth}
  \centering
  \caption[]{\small Conf-swap}
  \includegraphics[width=.99\textwidth]{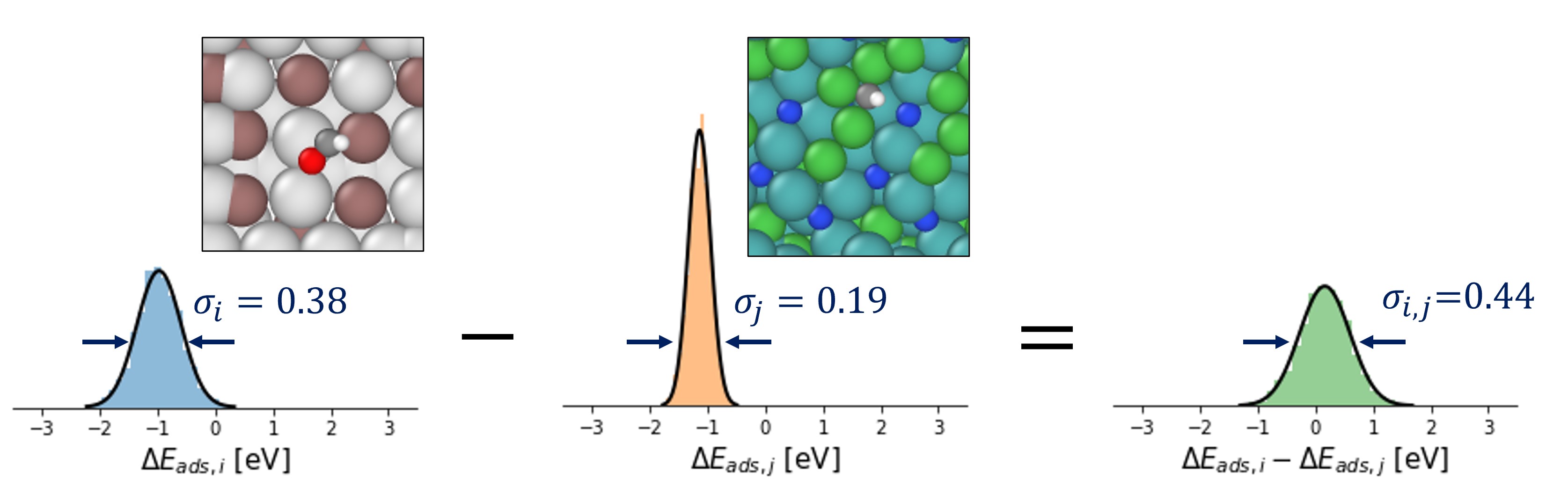}
  \label{fig:s1_b}
\end{subfigure}
\caption{Example cases of subtraction of energy ensembles. (a) All-swap (system \textit{i}: \ce{N4Ni8Mo12}/\ce{CH}, system \textit{j}: \ce{Sc4In2}/\ce{CHO}). (b) Conf-swap (system \textit{i}, \textit{j}: \ce{Pd4Sb8}/\ce{CH} with different configurations).}
\label{fig:s1}
\end{figure}

The error in the difference between two predicted values (X, Y) depends on the propagation of errors from the individual predictions, which can either partially cancel each other out or amplify the error. The propagated error ($\sigma_{\text{X, Y}}$) is determined by the individual error ($\sigma_{\text{X}}, \sigma_{\text{Y}}$) and their covariance (cov(X,Y)). Cancellation of individual errors can occur during subtraction when two predictions are correlated and have positive covariance. However, if the predictions are independent or have negative covariance, the error in the difference will be larger than the individual errors. Specifically, random error propagation is the term used to describe the case where the covariance is zero, assuming independent errors.

\begin{equation}
    \centering
    \sigma_{\text{X, Y}}^2 =\sigma_{\text{X}}^2 + \sigma_{\text{Y}}^2 - 2 \text{cov(X,Y)}
    \label{error_prob}
\end{equation}

This principle is also applicable to the subtraction of energy ensembles derived from BEEF-vdW calculations. When the energy ensembles of two pairs of systems are dissimilar and exhibit no positive correlation or covariance, the error distribution will be amplified after subtraction, as demonstrated in Figure \ref{fig:s1_a}. The resulting standard deviation of the energy difference ensemble is 0.44 eV, which is larger than the error propagation under the assumption of independent errors ($\sqrt{0.68^2 + 0.19^2}=0.42$ eV). Conversely, systems with similar electronic structures will show a reduced error distribution after subtraction, owing to the positive correlation and covariance between their energy ensembles, as illustrated in Figure \ref{fig:s1_b}. In such cases, the individual ensembles cancel out to some extent, resulting in a reduced standard deviation of the energy difference ensemble, namely, 0.17 eV, compared to the independent error propagation ($\sqrt{0.18^2 + 0.25^2}=0.31$ eV).

\newpage
\section{Error distributions of GNN predictions}
\label{sec:s2}

\begin{figure}[h!]
\centering
  \includegraphics[width=.76\linewidth]{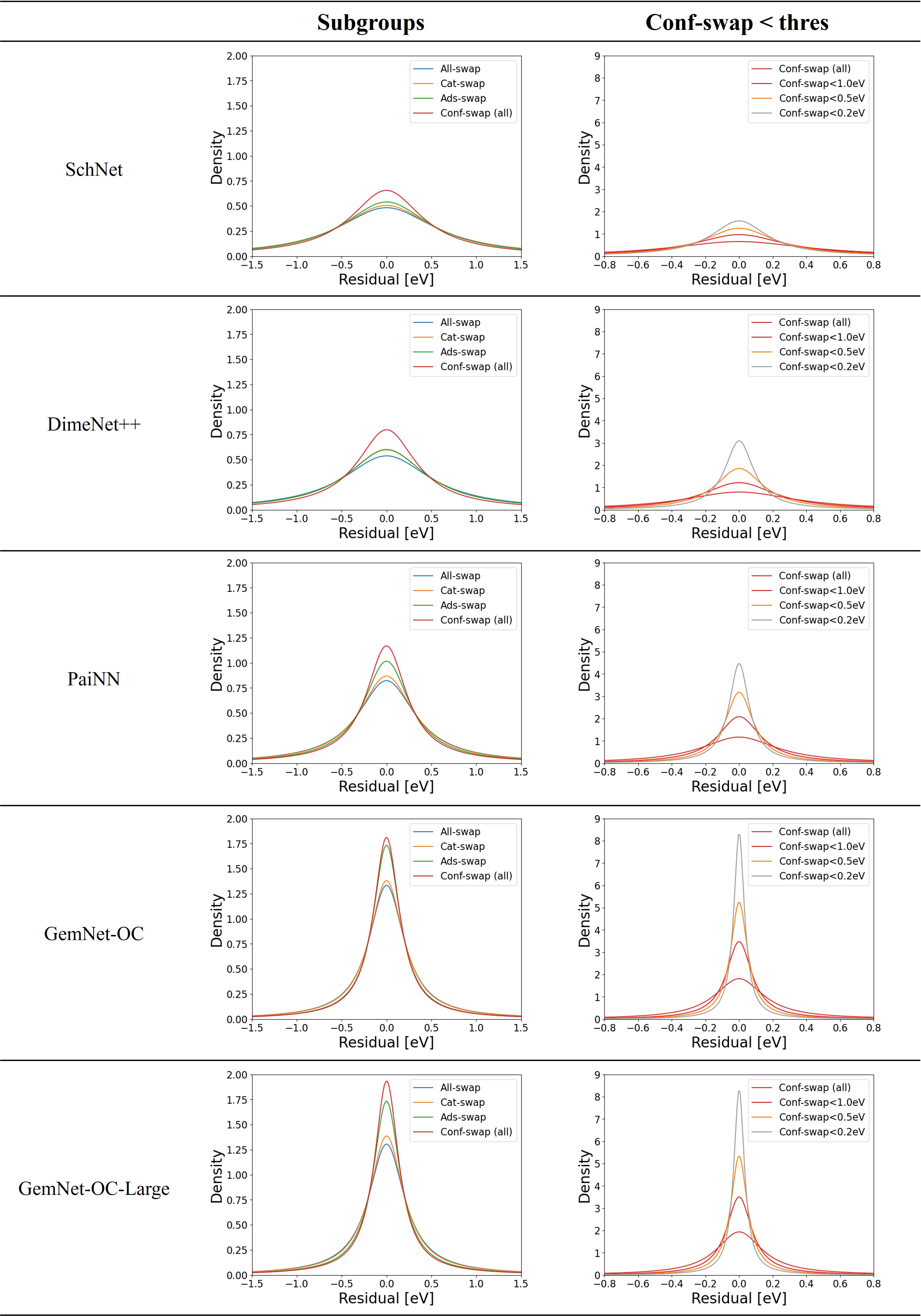}
\caption{Error distributions of energy difference predictions across subgroups and GNNs. The error distributions are obtained by calculating the residuals of GNN predictions ($\Delta \Delta E^{DFT} - \Delta \Delta E^{GNN}$). Narrowing down the subgroup results in a sharper error distribution for all GNNs. The error distribution becomes sharper across GNN models in the order of their accuracies: SchNet $<$ DimeNet++ $<$ PaiNN $<$ GemNet-OC $\simeq$ GemNet-OC-Large.}
\label{fig:s2}
\end{figure}


\newpage
\section{Contribution of ads-NN embeddings in energy prediction}
\label{sec:s3}

\begin{figure}[h]
\centering
  \includegraphics[width=.75\linewidth]{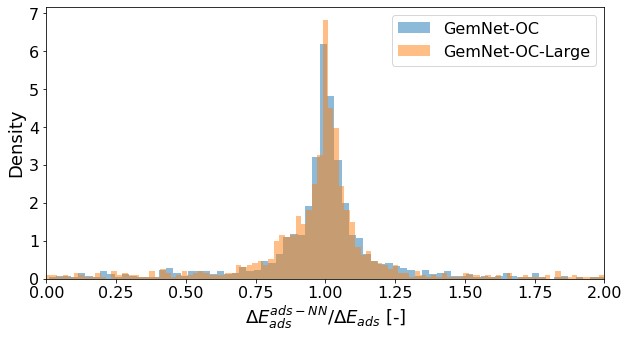}
\caption{Distribution of ratio between the energy predicted by the adsorbate atoms and their nearest neighbors on the catalytic surface (ads-NN) and the adsorption energy predicted by all atom nodes in the system.}
\label{fig:s3}
\end{figure}

The contribution of a specific set of atoms ($S$) to the energy prediction is quantified as $\left\lvert \sum_{i \in S}{e_i} / E \right\rvert$. In this study, $S$ corresponds to the atoms in ads-NN, and $E$ is the adsorption energy ($\Delta E_{\text{ads}}$). Data points clustered around 1 on the horizontal axis indicate a closer match between the energy predicted by ads-NN atoms and the actual adsorption energy. Notably, despite their small fraction of the systems, the ads-NN fraction accounts for 0.8 to 1.2 times the actual adsorption energy in approximately 76\% of cases, underscoring their significant role in predicting adsorption energy.

\newpage
\section{Embedding similarity analysis using the RV coefficient}
\label{sec:s4}

\begin{subequations}
\label{eq:RV_Coeff}
\begin{align}
\text{RV}(H_{i}^{n \times d}, H_{j}^{m \times d}) &= \frac{\text{COV}(H_{i}^{n \times d}, H_{j}^{m \times d})}{\sqrt{\text{VAR}(H_{i}^{n \times d}) \text{VAR}(H_{j}^{m \times d})}} \label{eq:rv} \\
\text{COV}(H_{i}^{n \times d}, H_{j}^{m \times d}) &= \sum_{l=1}^m \sum_{k=l+1}^n \text{cov}^2(h^k_i, h^l_j) \label{eq:cov} \\
\text{VAR}(H_i^{n \times d}) &= \sum_{l=1}^n \sum_{k=l+1}^n \text{cov}^2(h^k_i, h^l_i) \label{eq:var} 
\end{align}
\end{subequations}

To compare the similarity between two embedding matrices with different dimensionality, we use the RV coefficient, a multivariate generalization of the squared Pearson correlation coefficient. This coefficient enables comparisons between matrices of different sizes, as each catalyst system contains a distinct number of atoms in ads-NN, and measures the similarity of embedding matrices on a scale from 0 to 1. The RV coefficient normalizes the correlation between embedding matrices by the variance of each respective matrix, as shown in Equation \ref{eq:rv}. It provides a measure of the proximity of two sets of matrices in the latent space, and we calculate the average RV coefficient for pairs within a subgroup for comparison with other subgroups.